\documentclass[pra,aps,twocolumn,superscriptaddress,showpacs,reprint]{revtex4}
\usepackage{amsfonts}
\usepackage{amsmath}
\usepackage{amssymb}
\usepackage{graphicx}
\usepackage{epstopdf}
\usepackage{subfigure}

\DeclareMathOperator{\sech}{sech}
\providecommand{\U}[1]{\protect\rule{.1in}{.1in}}

\newcommand{\subfigimg}[3][,]{%
  \setbox1=\hbox{\includegraphics[#1]{#3}}
  \leavevmode\rlap{\usebox1}
  \rlap{\hspace*{1pt}\raisebox{\dimexpr\ht1-2\baselineskip}{#2}}
  \phantom{\usebox1}
}

\begin{document}

\preprint{HEP/123-qed}
\title[Generation of nonlinear vortex precursors]{Generation of nonlinear vortex precursors}
\author{Yue-Yue Chen}
\affiliation{State Key Laboratory of High Field Laser Physics, Shanghai Institute of
Optics and Fine Mechanics, Chinese Academy of Sciences, Shanghai 201800,
China}
\affiliation{University of Chinese Academy of Sciences, Beijing 100039, China}
\author{Xun-Li Feng}
\affiliation{Department of Physics, Shanghai Normal University, Shanghai 200234, China}
\author{Chengpu Liu}
\email{chpliu@siom.ac.cn}
\affiliation{State Key Laboratory of High Field Laser Physics, Shanghai Institute of
Optics and Fine Mechanics, Chinese Academy of Sciences, Shanghai 201800,
China}

\begin{abstract}
We numerically study the propagation of a few-cycle pulse carrying orbital angular momentum (OAM) through a dense atomic system. Nonlinear precursors consisting of high-order vortex harmonics are generated in the transmitted field due to ultrafast Bloch oscillation. The nonlinear precursors survive to propagation effects and are well separated with the main pulse, which provide a straightforward way of measuring precursors. By the virtue of carrying high-order OAM, the obtained
vortex precursors as information carriers have potential applications in
optical information and communication fields where controllable loss, large
information-carrying capacity and high speed communication are required.
\end{abstract}

\volumeyear{year}
\volumenumber{number}
\issuenumber{number}
\eid{identifier}
\date[Date text]{date}
\received[Received text]{date}
\revised[Revised text]{date}
\accepted[Accepted text]{date}
\published[Published text]{date}
\startpage{1}
\endpage{102}
\maketitle

More than one century ago, the concept of optical precursors \cite{precursor}
emerged from the seminal works of Sommerfeld \cite{Sommerfeld} and Brillouin
\cite{Brillouin,Brillouin2} on asymptotic description of ultrawideband
dispersive pulse propagation in linear dielectrics. This description is
derived from the exact Fourier-Laplace integral representation of the
propagated linearly polarized plane wave field \cite%
{Oughstun} and approximating the material system as a
collection of linear oscillators \cite{Lorentz}.
To observe precursors, considerable theoretical \cite%
{Cartwright,Jeong,Du,Macke,LeFew} and experimental \cite%
{Aaviksoo,Choi,HJ,SD} studies on optical precursors have been done over
years. However, many initial works focused on an
opaque medium with single or multiple Lorentz absorption lines, where the
main signal is either absorbed or cannot be well separated from precursors.
This provokes controversies about the existence of precursors
\cite{Alfano,osterberg}.

Recently, it has been reported that the precursors also exist in nonlinear
interactions regime \cite{Ding,Marskar,Palombini}. Ding \textit{et al.}
observed optical precursors, which consist of the high-frequency components
of the signal pulse, in four-wave mixing based on a cold-atom gas \cite{Ding}.
Palombini \textit{et al.} studied the effects of a nonlinear medium
response on precursor formation using the split-step Fourier method \cite%
{Palombini}. In both cases, the nonlinear precursors are obtained with the
main pulse absorbed by the medium. A scheme to obtain nonlinear precursors
in a two-level system with rotating-wave approximation (RWA) and slowly
varying envelope approximations (SVEA) is proposed, where the precursors are
well separated with main pulse \cite{Marskar}. However, for the few-cycle
physics, the standard approximations used in traditional nonlinear optics
are no longer appropriate \cite{Rothenberg}. Thus, the full wave
Maxwell-Bloch (MB) equations without SVEA and RWA need to be solved.

On the other hand, light beams can exhibit helical wave fronts \cite{Yao,Arnold}.
High-order optical vortex beam has many
potential applications such as generating Multi-dimensional entanglement to
support efficient use of communication channels in quantum cryptography \cite%
{Mair}, and photoexciting atomic levels without the restriction of standard
dipolar selection rules \cite{Afanasev,Picon}. Recently, schemes using gas
\cite{Matsko} and plasma \cite{ZX} as mode converters to increase the OAM of
a beam charge have been proposed. Their schemes conflate high-order
harmonics generation (HHG) and OAM, imprinting the phase twist to the
fundamental field instead of the short-wavelength radiation directly.
However, these schemes either require extremely intense ($\sim 10^{15}$W$/$cm$^{2}$%
) or even relativistic laser pulses ($\sim 10^{22}$W$/$cm$^{2}$), where
damage thresholds of the nonlinear medium have to be considered,
or are restricted to an ultrathin medium ($\sim 1 \mu$m), where propagation effect are ignored.

In this paper, we presents a scheme to generate nonlinear precursors consisting of high-order vortex harmonics
in relatively low energy physics, where the propagation effects are considered. A dense two-level atomic medium is used as
a mode converter to manipulate the OAM of a few-cycle helical pulse.
Interestingly, nonlinear precursors consisting of high-order vortex harmonics appear in the front of the fundamental field. Compared with that obtained in reflected
field \cite{ZX}, the high-order vortex harmonics existed in the precursors
are instinctively separated from the fundamental mode, sparing the necessity
of a filter to observe the harmonics.
Therefore, our proposal generates high-order vortex harmonics with
merits of precursors, relaxes the requirement of extremely intense pulse for
ionization or plasma generation, and takes propagation effects into consideration. Using high-order vortex
precursors as information carriers in quantum information can favor the
realization of high-speed communication, enhance the efficiency of
communication channels, and improve the robustness to resonant absorption
loss.

A few-cycle Laguerre-Gaussian (LG) laser pulse propagates along $z$ in
vacuum and is incident on a dense two-level atomic medium. Assuming the
driving pulse is linearly polarized along $x$, the incident pulse takes the form $E(t=0,z)=E_{lp}\cos [\omega _{p}(z-z_{0})/c]%
\sech[1.76(z-z_{0})/(c\tau _{p})]\hat{e}_{x},$ where $\omega _{p}$ is the
carrier frequency, $\tau _{p}$ the full width at half maximum of the
pulse intensity envelop, $\hat{e}_{x}$ the unit vector in $x$ direction. The
initial position $z_{0}$\ is set to be $3\mu $m to avoid the pulse
penetrating into the medium at $t=0$. The amplitude $E_{lp}$ is defined by
\cite{Allen}

\begin{eqnarray}
E_{lp}(t &=&0,z)=\frac{E_{0}}{(1+\tilde{z}^{2}/z_{R}^{2})^{1/2}}(\frac{r}{a(%
\tilde{z})})^{l}L_{p}^{l}(\frac{2r^{2}}{a^{2}(\tilde{z})})  \notag \\
&&\times \exp (-\frac{r^{2}}{a^{2}(\tilde{z})})\exp (-\frac{ikr^{2}\tilde{z}%
}{2(\tilde{z}^{2}+z_{R}^{2})})\exp (-il\phi )  \notag \\
&&\times \exp (i(2p+l+1)\tan ^{-1}\frac{\tilde{z}}{z_{R}}),
\end{eqnarray}%
where $\tilde{z}=z-z_{0},$ $E_{0}$\ is the peak amplitude of the incident
pulse, $z_{R}$ the Rayleigh range, $a(\tilde{z})$ the radius of the beam, $%
L_{p}^{l}$\ associated Laguerre polynomial, and the beam waist $a_{0}$ is at
$z=z_{0}$. The characteristic helical phase profiles of optical vortices are
described by $\exp (-il\phi )$ multipliers, where $l(l=0,\pm 1,\pm 2...)$ is
the topological charge corresponding to the mode order and $\phi $ the
azimuthal coordinate. The integer $p$ denotes the number of radial nodes in
the mode profile. The three-dimensional Maxwell's equations in an isotropic
medium take the form

\begin{eqnarray}
\frac{\partial \mathbf{H}}{\partial t} &=&-\frac{1}{\mu _{0}}\nabla \times
\mathbf{E},  \notag \\
\frac{\partial \mathbf{E}}{\partial t} &=&\frac{1}{\epsilon _{0}}\nabla
\times \mathbf{H}-\frac{1}{\epsilon _{0}}\frac{\partial \mathbf{P}}{\partial
t}.
\end{eqnarray}%
The macroscopic polarization induced by the linearly polarized electric
field is $P_{x}\hat{e}_{x}.$ $P_{x}=Ndu$ is
associate with the off-diagonal density-matrix element $\rho
_{12}=(u+i\upsilon )/2,$ $N$ the density and $d$ the dipole moment. The
population inversion between the excited 2 and the ground state 1 is denoted
by $w=\rho _{22}-\rho _{11}$. $u,\upsilon $ and $w$ obey the following set
of Bloch equations,

\begin{eqnarray}
\frac{\partial u}{\partial t} &=&-\gamma _{2}u-\omega _{0}v,  \notag \\
\frac{\partial v}{\partial t} &=&-\gamma _{2}v+\omega _{0}u+2\Omega w,
\notag \\
\frac{\partial w}{\partial t} &=&-\gamma _{1}(w-w_{0})-2\Omega v.
\end{eqnarray}%
Where $\gamma _{1},\gamma _{2}$ are, respectively, the population and
polarization relaxation rate, $\omega _{0}$ the resonant frequency, $\Omega
(z,t)$ the Rabi frequency, and $w_{0}$
is the initial population difference.

The full wave MB equation can be solved by adopting Yee's finite-difference
time-domain (FDTD) discretization method for the electromagnetic fields \cite%
{Yee,Taflove} and the predictor-corrector method \cite{Ziolkowski,Hughes} or
the fourth order Runge-Kutta method \cite{Yu} for the medium variables. The
medium is initialized with $u=\upsilon =0,w_{0}=-1.$ The following
parameters are used to integrate the MB equations: $\omega _{0}=\omega
_{p}=2.3$fs$^{-1},$ $d=2\times 10^{29}$A$\ast $s$\ast $m, $\gamma
_{1}^{-1}=1 $ps$,\gamma _{2}^{-1}=0.5$ps$,\tau _{p}=5$fs$,a_{0}=7\mu $m,
medium length $L=25\mu $m, $\Omega _{0}=1.408$fs$^{-1},$ the corresponding
pulse area is $A(z)=d/\hbar \int_{-\infty }^{\infty }E_{0}(z,t^{^{\prime
}})dt^{^{\prime }}=\Omega _{0}\tau _{p}\pi /1.76=4\pi $ \cite{Kalosha}.
Defining a collective frequency parameter $\omega _{c}=Nd^{2}/\epsilon
_{0}\hbar =0.1$fs$^{-1}$ to present the coupling strength between medium and
field. Our simulation region is padded with perfectly matched layers that
prevents back reflection from the truncated simulation region.

\begin{figure}[b]
\begin{center}
\includegraphics[width=
0.5\textwidth]{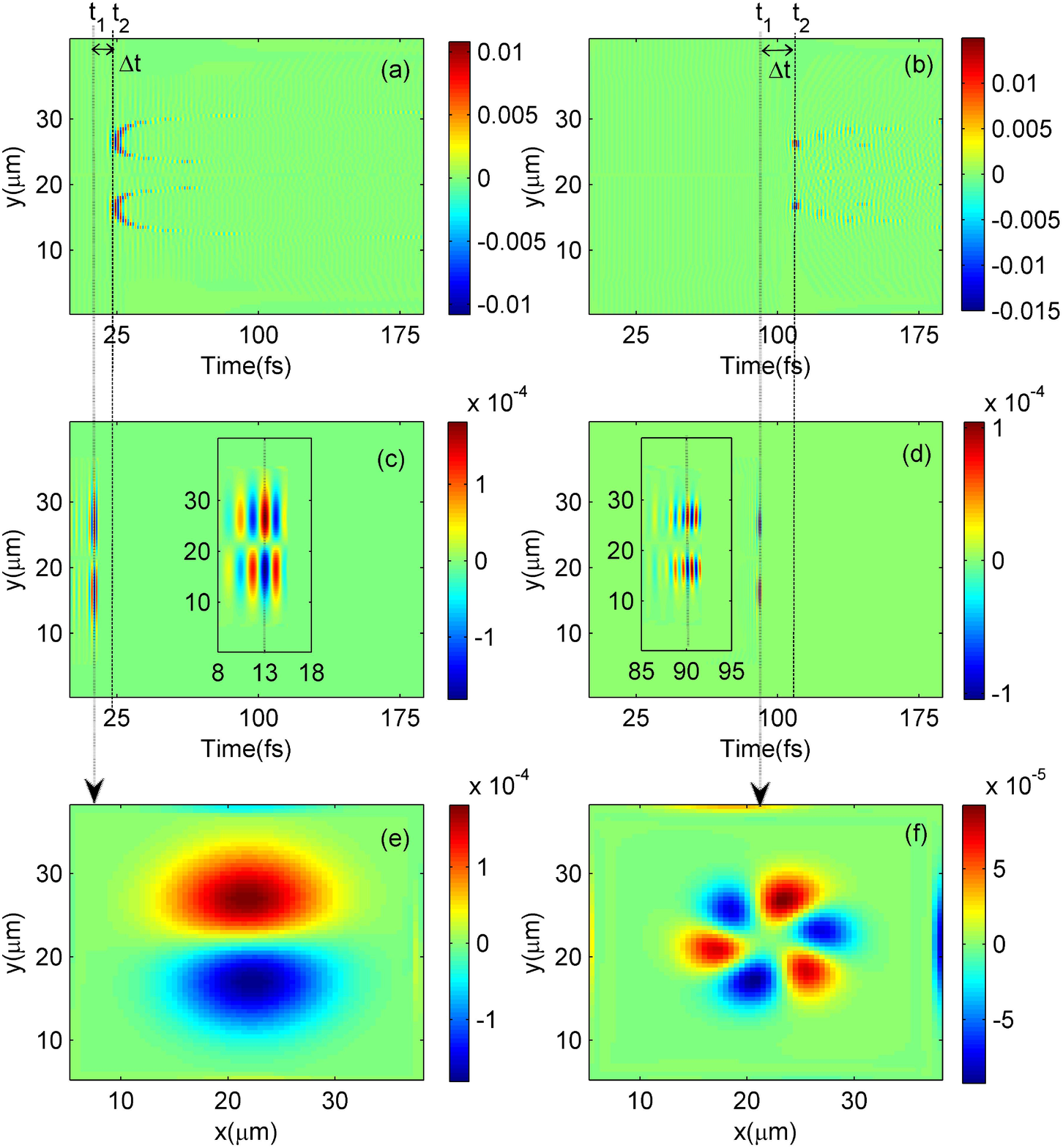}
\end{center}
\begin{flushleft}
\caption{(color online) (top) Time evolutions of $E_{x}$ at $z_{1}=9\protect\mu $%
m (left) and $z_{2}=30\protect\mu $m (right). (middle) The corresponding evolutions of
the leading part of the precursors and their enlarged view in the inserts.
(bottom) The
corresponding transverse distribution of $E_{x}$ at $t_{1}$. }
\end{flushleft}
\end{figure}

With the given parameters, the evolutions of a LG$_{10}$ beam at $%
z=9$ $\mu $m and $30$ $\mu $m are obtained, as shown in Fig. 1. The
evolution of LG$_{10}$ beam with two lobes is quite similar with that of two
out-of-phase $2\pi $\ pulses \cite{Niu}. Since the local delays are
inversely proportional to the local Rabi frequency, each lobe of the LG$%
_{10} $ beam evolves progressively into a crescent-shaped pulse,
accompanying with self-focus caused by the diffraction-induced inward flow
of energy from the outer rings \cite{Xia,Lamare}, as shown in Fig. 1(a).
However, due to transverse effect, the crescent-shaped fields are unstable
\cite{Xia}. Each of them breaks up into a leading $2\pi $ SIT soliton
located at lobe-peak and several small-area fragments at wings. As the pulse
further propagates, only the two leading $2\pi $ SIT-like solitons survived. All
the small-area fragments experience energy decrease and vanish at a large
distance, as shown in Fig. 1(b).

\begin{figure}
  \begin{center}
  \begin{tabular}{@{}p{0.5\linewidth}@{\hspace{0.05cm}}p{0.5\linewidth}@{}}
    \subfigimg[width=\linewidth]{}{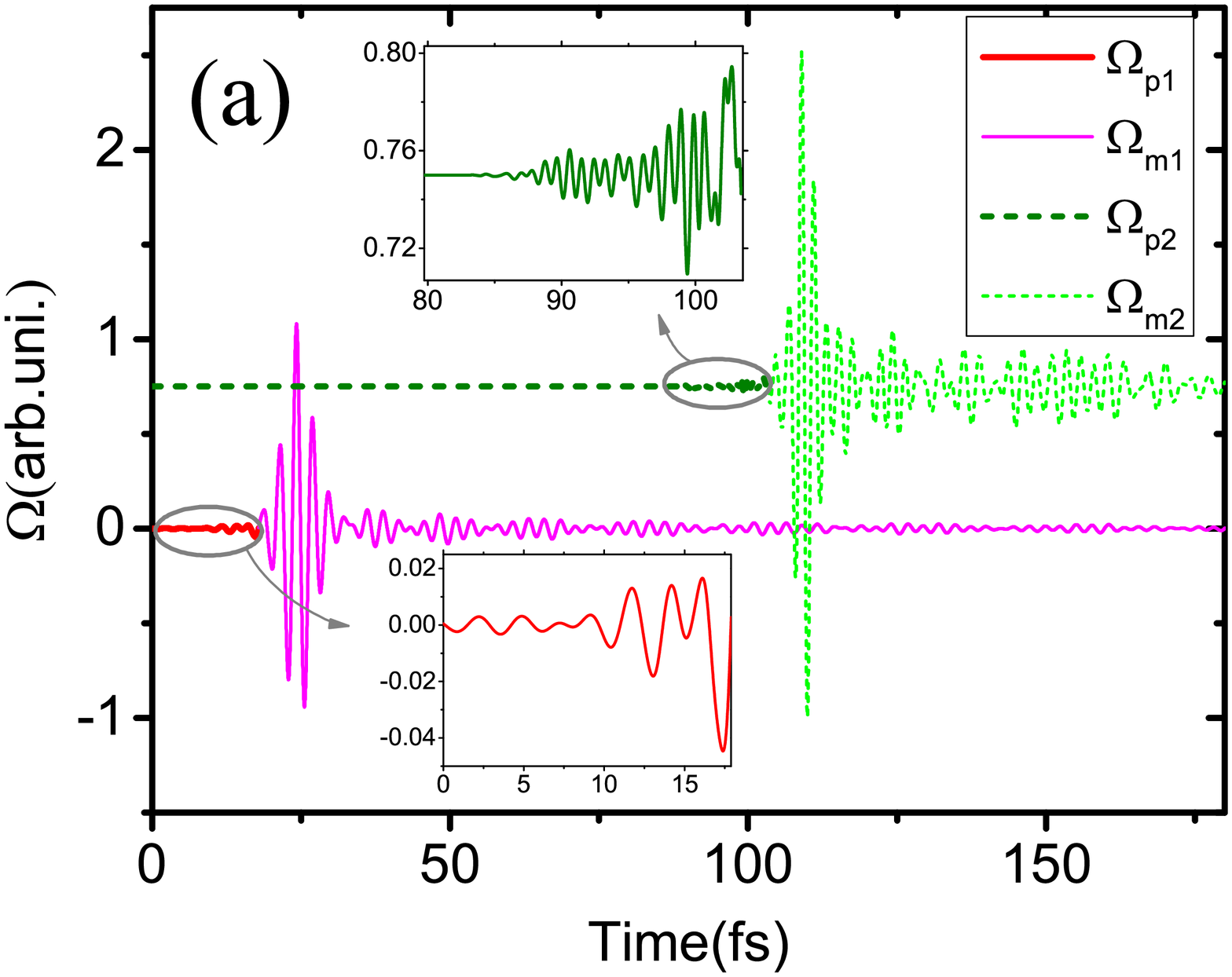} &
        \subfigimg[width=\linewidth]{}{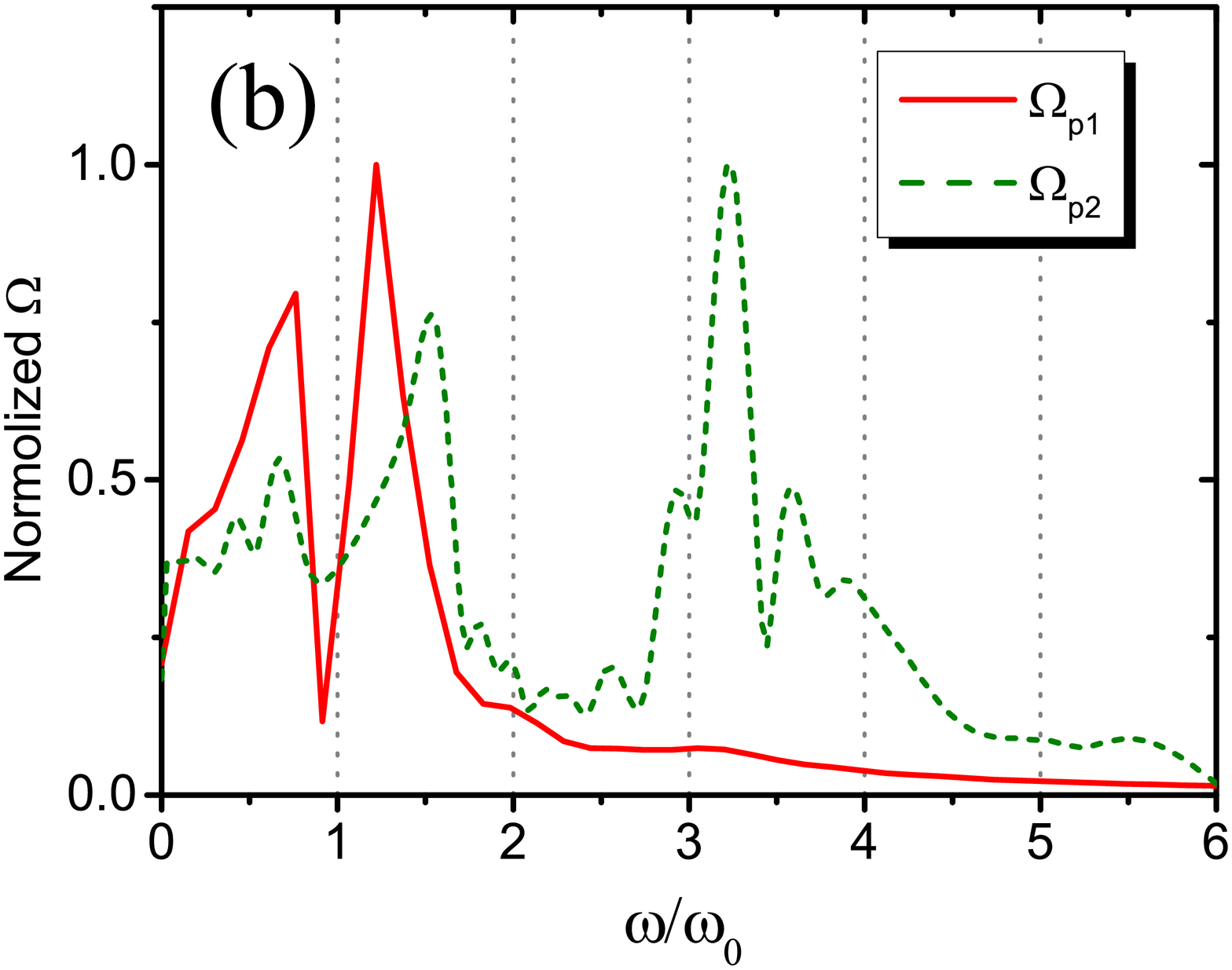}\\
    \subfigimg[width=\linewidth]{\quad\quad(c)}{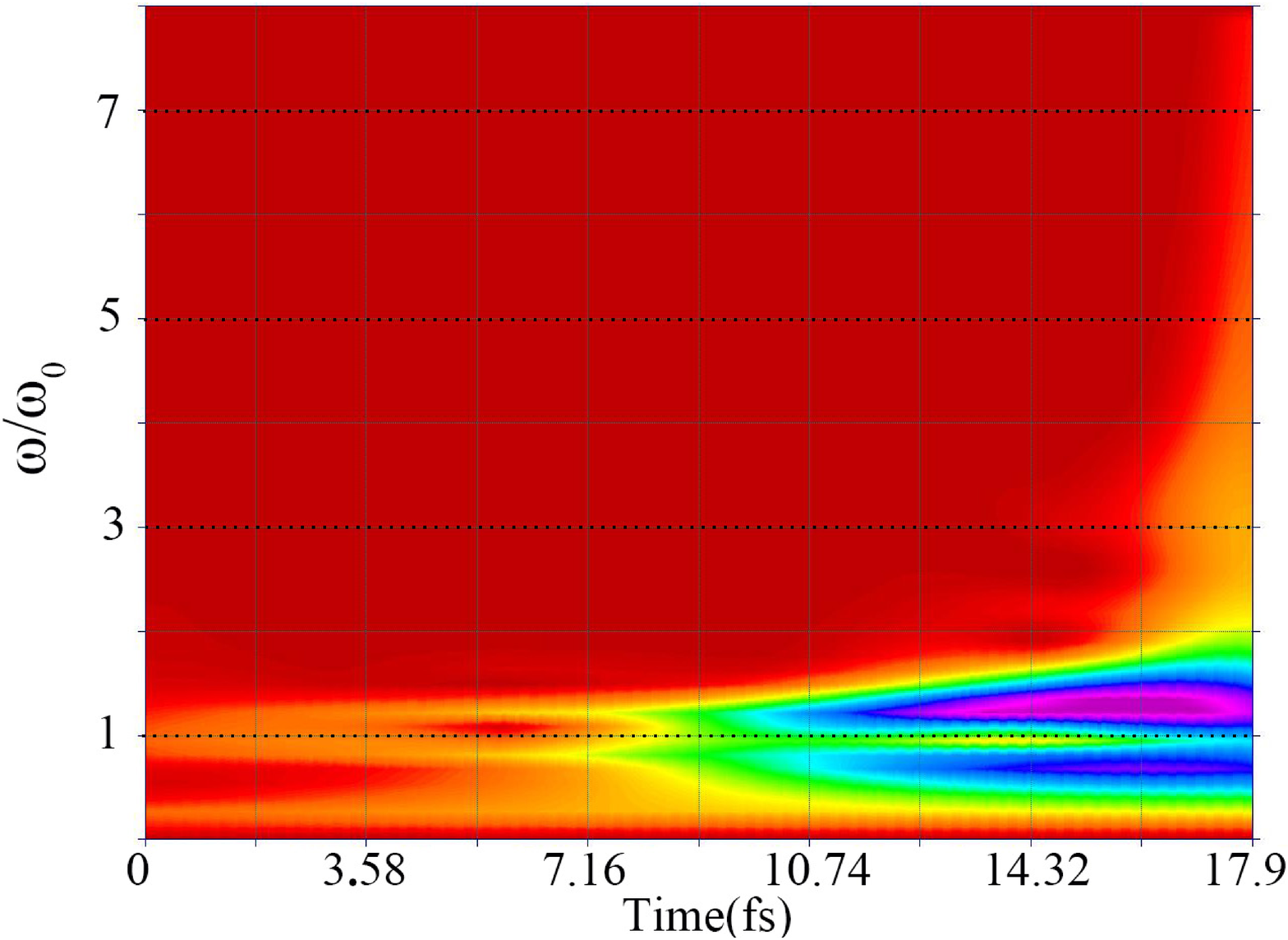} &
    \subfigimg[width=\linewidth]{\quad\quad(d)}{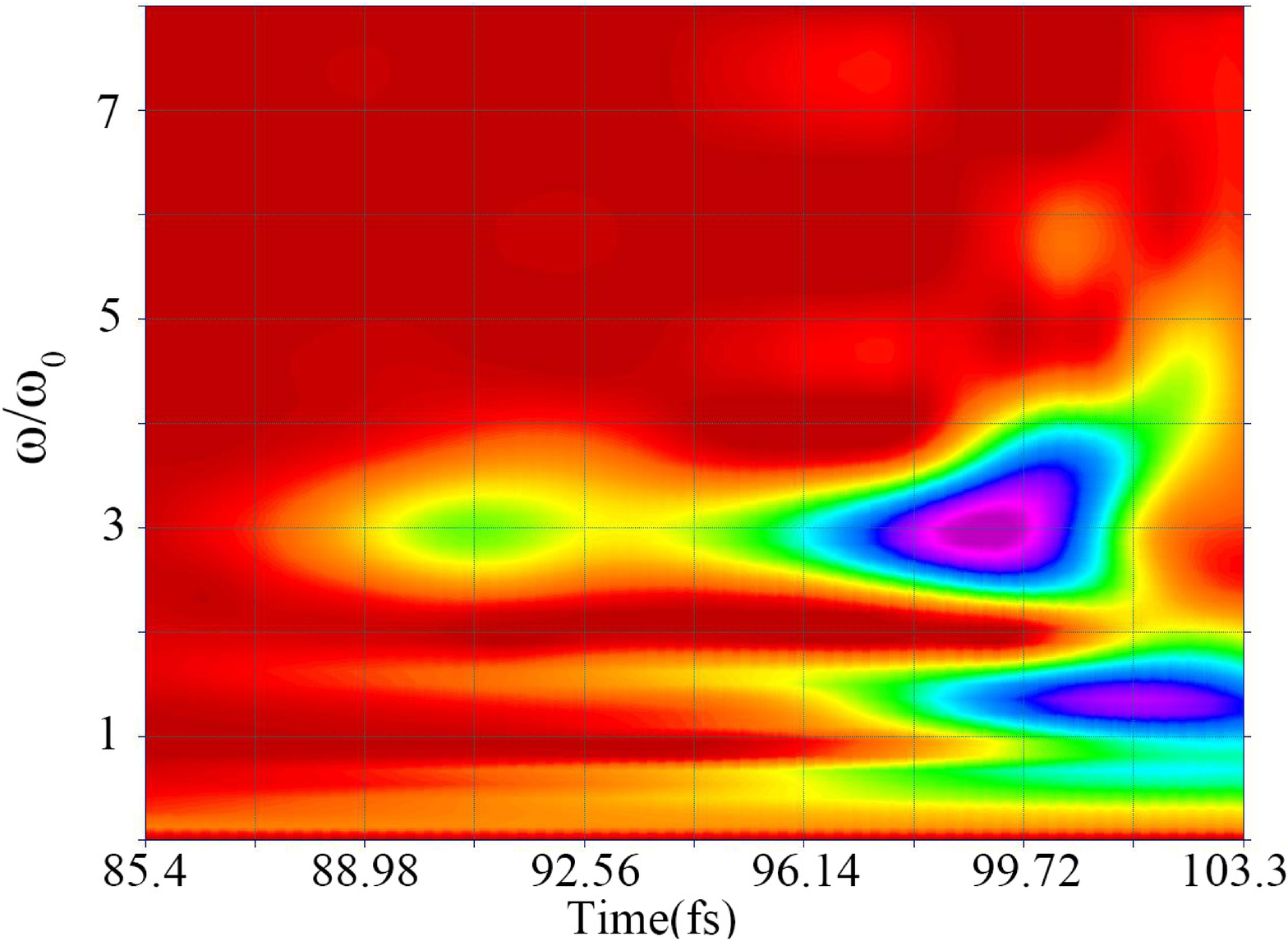}
  \end{tabular}
  \end{center}
  \caption{(color online) The precursors (a) and their spectra (d)
  correspond to $z_{1}=9\protect\mu $m (solid line) and $z_{2}=30\protect\mu $m (dashed line), respectively, at $x=21.5\mu$m and $y=16.5\mu$m.
  The time-frequency analysis graphs of precursors
correspond to $z_{1}$ (c) and $z_{2}$ (d), respectively.}
\end{figure}

More importantly, precursors ahead of the
main pulse are generated during the propagation.
The precursors consist of off-resonant frequency components
and propagate with the speed of light in vacuum, while
the main pulse is delayed by the resonant medium.
The leading part of the precursors are shown in Figs. 1(c) and 1(d). At the beginning of the propagation ($z=9\mu$m), the electric field of precursors in longitudinal plan is similar with that of incident pulse, in terms of the beam diameter and carrier frequency, as shown in Fig. 1(c). In addition, the transverse plane of the precursors in Fig. 1(e) shows the LG$_{10}$-like mode as the driving pulse. Later, at $z=30\mu$m,
the precursors have a much smaller beam diameter and a much higher carrier frequency than that at the beginning. The decrease of beam diameter is caused by self-focusing. The corresponding transverse electric field in Fig. 1(f) shows a LG$_{30}$-like mode. Meanwhile, the distance between the leading of the precursors and the main pulses, i.e. $\Delta t=t_{2}-t_{1}$, increases along the propagation path, which confirms the theory that precursors propagate faster than main pulse.

To have a overall understanding of precursors, the electric fields at $z_{1}$ and $z_{2}$ are obtained. As shown in
Fig. 2(a), precursors (thick line) emerge in front of the main pulse (thin line) and evolve to a structure with fast oscillations. The spectra in Fig. 2(b) show that dominate frequency components of the precursors change from near-resonant frequency to third harmonic. To further investigate the frequency changes of precursors induced by propagation, we performed time-frequency analysis to $z_{1}$ and $z_{2}$, respectively.
For $z_{1}$, where the interaction length is relatively short, the precursors are linear. These linear precursors contain high- and low-frequency components corresponding to Sommerfeld and Brillouin precursors, respectively, as shown in Fig. 2(c). Then, odd harmonics are generated but barely separated with main pulse.
Since the atoms cannot respond quickly to rapid changes in electric field corresponding to odd harmonics, the high-frequency components
pass through the atoms without any delay and are gradually separated from the delayed main pulse. As shown in Fig. 2(c), for $z_{2}=30\mu$m, the third harmonic is well separated with main pulse and become the dominant frequency components of precursors. In this case, the linear precursors becomes nonlinear. Thus, the transformation of linear to nonlinear precursors occurs along propagation path.

\begin{figure}[b]
\begin{center}
\includegraphics[width=
0.5\textwidth]{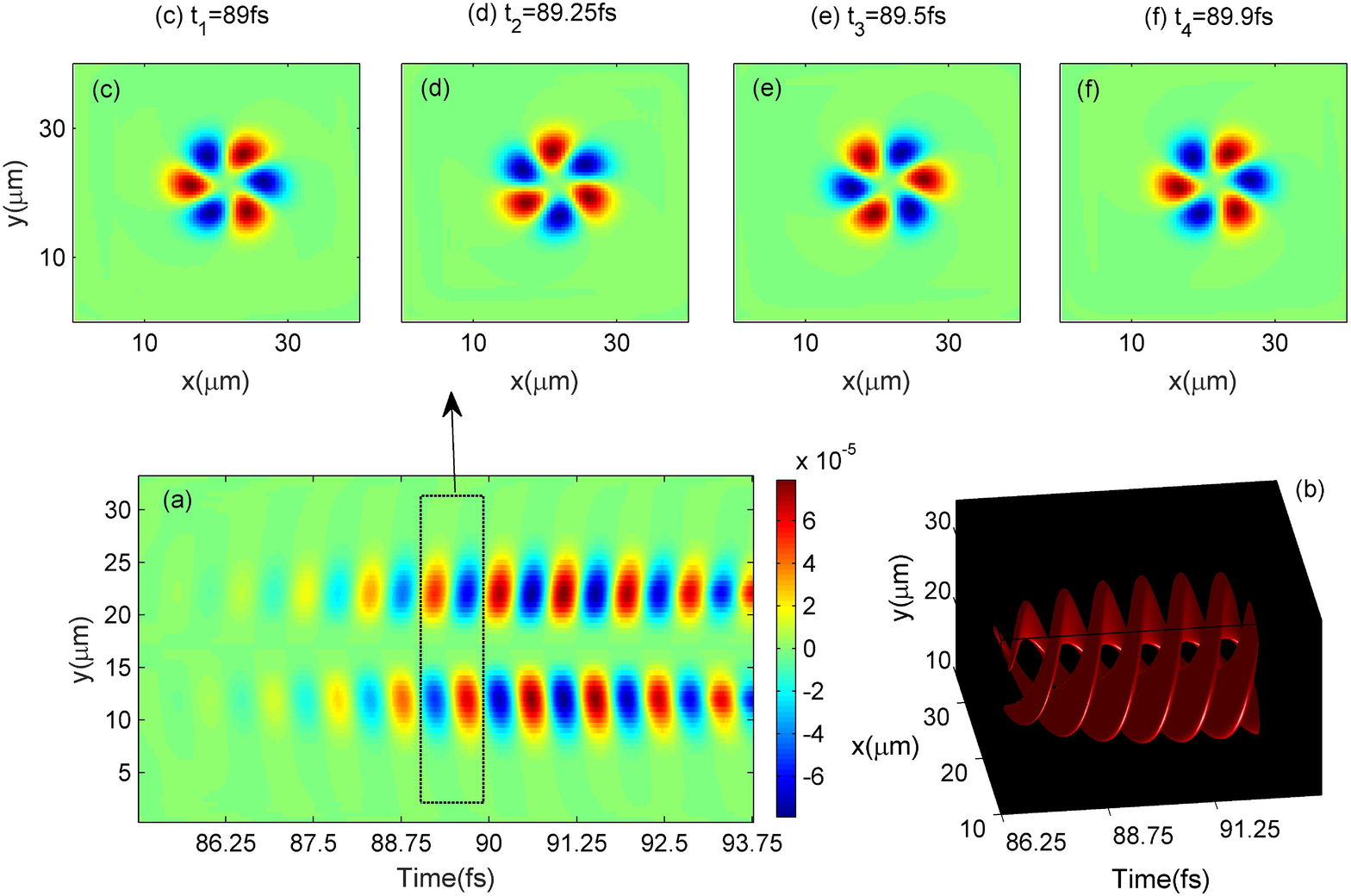}
\end{center}
\caption{(color online) (a) The time evolution of $E_{x}$ at $z_{3}=30%
\protect\mu $m during [85fs, 93.75fs]. (b) The corresponding 3D spatial
distribution of $E_{x}$. (c-f) The distribution of $E_{x}$ in $x-y$ plan
during the time of rotating one loop.}
\end{figure}
Since the nonlinear precursors are separated from the main pulse by nature,
their vortex information can be obtained without the necessary of filtering
the fundamental field. It can be seen from Figs.
3(e)-3(f) that the changes of the electric field distributions within one loop
show clearly helical feature. The distance of rotating one
loop of $E_{x}$ is approximately $\lambda _{3}=0.27\mu $m, which is equal to
the wavelength of the third harmonics $\lambda _{0}/3$. The corresponding
three-dimensional temporal evolution of the field in Fig. 3(a) is shown in
Fig. 3(b), where the helical structure of third harmonic field can be seen clearly. Thus,
third-order vortex precursors appear at the front of the main pulse.

\begin{figure}[t]
\begin{center}
\includegraphics[width=
0.5\textwidth]{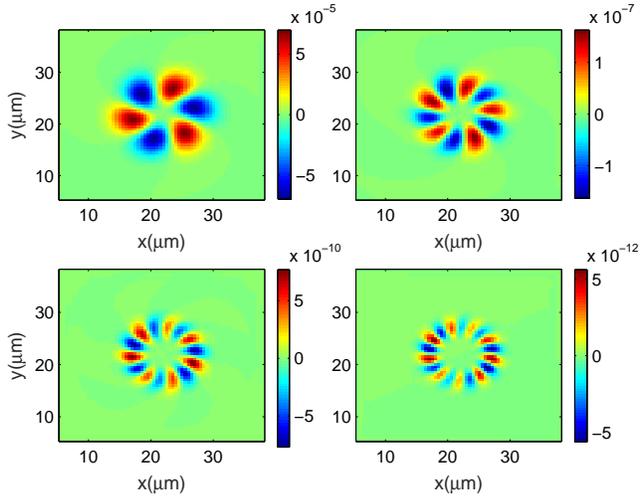}
\end{center}
\caption{(color online) Electric field distribution of (a) third, (b) fifth,
(c) seventh, and (d) ninth harmonics in $x-y$ plane at $t_{1}=90fs$ and $%
z_{3}=30\protect\mu $m.}
\end{figure}

\begin{figure}[b]
\begin{center}
\includegraphics[width=
0.5\textwidth]{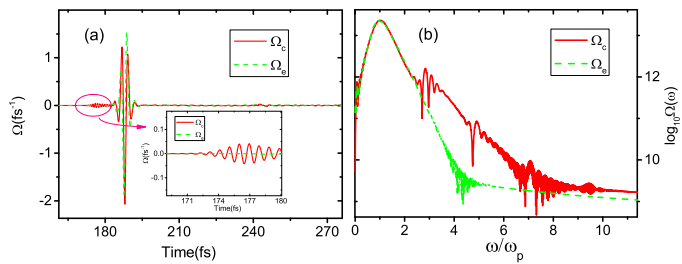}
\end{center}
\caption{(color online) The influence of RWA to transmitted fields (a) and spectra (b). The solid and the dashed lines correspond to the case
without and with RWA, respectively. The inserts in (a) is the enlarged view of the precursors.
}
\end{figure}
Moreover, as showed in Fig. 2 (d), higher-order harmonics coexist with third harmonic in nonlinear precursors. The transverse distributions of the third, fifth, seventh and ninth harmonics of precursors at $z_{3}$ and $t_{1}$ are shown in Fig. 4. According to the number of interwind
helices, the azimuthal modes of these harmonics are $l=3,5,7,9$,
respectively. The topological charge of the $l$th-order harmonic is $l$,
which is expected from the HHG theory \cite{Garcia,Patchkovskii}. Since the
intensity of the $l$th-order harmonics is inversely proportional to $l$, the
transverse structure of precursors is dominated by the LG$_{30}$-like
mode, as shown in Fig. 1(f). Thus, nonlinear precursors consisting of high-order
vortex harmonics appear in the
front of the main pulse with a distance that increases during the
propagation.

Finally, the root of the nonlinear precursors is discussed based on one-dimensional MB equations with and
without RWA. As shown in Fig. 5, in the case without RWA,
precursors appear in the front of the 2$\pi$ SIT soliton, and
odd harmonics corresponding to nonlinear precursors emerge in the transmitted spectrum,
which are consistent with the above description of precursors in three-dimensional case. However, in the framework of RWA, both the precursors and the odd-order harmonics disappear. This is because, the carrier induces rapid changes of the
refractive index, which in turn can reshape the electric field. When the
change of refractive index is proportional to the square of the
electric-field value, the two-level atom has a Kerr-medium characteristic
\cite{Ziolkowski}. The Kerr nonlinearity, such as intrapulse four-wave
mixing and self-phase modulation, could contribute to the occurrence of
odd harmonics. The generated odd harmonics propagate faster than the main pulse,
forming the preceding nonlinear precursors \cite{Ding}. While in the framework of RWA,
the refractive index barely changes and the nonlinear effect
corresponding to ultrafast Bloch oscillation is ignored, which is responsible for
the missing of nonlinear precursors in the transmitted spectrum. Therefore, the nonlinear precursors in the
transmitted spectrum is generated by carrier effects associated with
ultrafast Bloch oscillation.

Our scheme provides a way to manipulate the OAM of a beam by means of
nonlinear optics without the requirement of extremely intense light. It also
combines the merits of both precursors and high-order vortex harmonics. On
one hand, thanks to the advantage of precursors, the obtained high-order
vortex harmonics can propagate with the speed of light in vacuum, and
instinctively separate from the intense fundamental mode. The obtained
high-order vortex precursors can survive the propagation and are resilient
to absorption. In addition to be information carriers used in high speed
communication, precursors can also be used as optical probes of biological
tissues or in underwater communication. On the other hand, since the
precursors consist of high-order vortex harmonics, they have potential
application in quantum information and communication where multi-dimensional
entanglement are required.

\bigskip

This work is supported by the National Natural Science Foundation of China
(NNSF, Grant No.11374318 and No. 11374315). C.L. is appreciated to the
supports from the 100-Talents Project of Chinese Academy of Sciences and
Department of Human Resources and Social Security of China.

\end{document}